\documentclass{article}

\usepackage{WEJCF}

\usepackage[utf8]{inputenc} 
\usepackage[T1]{fontenc}    
\usepackage{siunitx}
\usepackage{hyperref}       
\usepackage{url}            
\usepackage{booktabs}       
\usepackage{lipsum}
\usepackage{ucs}
\usepackage{amsmath}
\usepackage{amsfonts}
\usepackage{amssymb}
\usepackage{graphicx}
\usepackage{setspace}
\usepackage{rotating}
\usepackage{multirow}
\usepackage{mathrsfs}
\usepackage{subfig}
\usepackage{bigstrut}
\usepackage{pdfpages}

\begin{document}



\title{Production of Jets at STAR}

\author{ Michal Svoboda \\ \textit{on behalf of the STAR Collaboration}}

\maketitle

\begin{center}
    \textbf{Abstract}
\end{center} 

Jets serve as an important tool to probe QCD both in the vacuum and in the hot and dense medium. The STAR experiment at RHIC plays a key role in studying QCD phenomena across different collision systems ($p$+$p$, $p$+A, A+A), offering access to a kinematic regime that complements that of the LHC. Building on recent jet and event activity studies at STAR, we present recent measurements on charged-particle jets at $\sqrt{s_{\mathrm{NN}}}~=~200$ GeV. In $p$+Au collisions, we explore event activity (EA) measured in the Au-going direction and its correlation with particle production at mid-rapidity. While soft particle production increases with EA, high-$p_{\mathrm{T}}$ jets are found to be inversely related to EA. Ratios of $p_{\mathrm{T}}$ imbalance and azimuthal dijet separation between high- and low-EA events show no significant differences, suggesting no strong evidence of jet quenching in high-EA $p$+Au collisions. In Au+Au collisions, we report semi-inclusive measurements of jets recoiling from $\gamma$ and $\pi^0$ triggers, using mixed-event techniques to subtract background and study jet suppression, intra-jet broadening, and acoplanarity. Additionally, we present inclusive charged-particle jet spectra corrected for background fluctuations, extending the kinematic reach of previous measurements. These results provide crucial insight into the modification of jets in the medium and contribute to a deeper understanding of QCD in heavy-ion collisions.

\section{Introduction}

Shortly after the Big Bang, the Universe existed in a unique state of matter known as the Quark-Gluon Plasma (QGP), where quarks and gluons are not confined within hadrons but formed a hot, dense medium. This state of matter can be recreated for a brief moment in high-energy heavy-ion collisions at large-scale particle colliders such as the Large Hadron Collider (LHC) at CERN and the Relativistic Heavy Ion Collider (RHIC) at BNL in the USA.

One of the primary tools to study the QGP is the measurement of jets—collimated sprays of particles resulting from the fragmentation and hadronization of hard-scattered partons. Jets are produced in both proton-proton ($p+p$) and heavy-ion collisions, making them an excellent probe for exploring the properties of the QGP. By comparing jet properties between $p+p$ and heavy-ion collisions, the modifications induced by the medium can be studied.

A key observable in this context is jet quenching, which refers to the energy loss experienced by high-energy partons as they traverse the QGP. Jet quenching manifests itself through the suppression of high transverse momentum particle yields and modifications to the structure of fully reconstructed jets, providing crucial insight into the interactions between hard probes and the QGP medium.


\section{Experimental Setup}

\begin{figure}[h]

\centering
\includegraphics[width=10 cm]{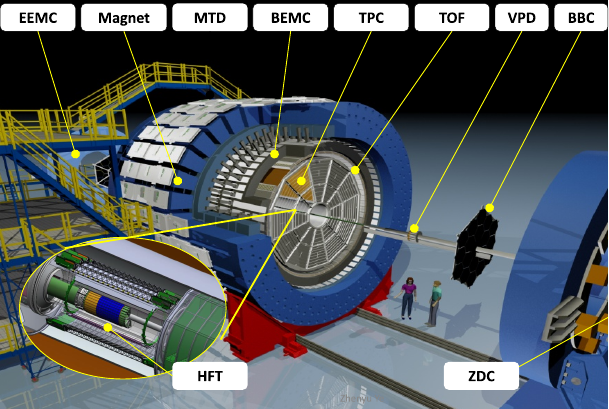}
\caption{STAR detector and its sub-detectors. Taken from Ref. \cite{Tannenbaum2019}\label{STAR}.}
\end{figure}

The RHIC is an exceptionally versatile accelerator capable of delivering beams of various species, from protons to uranium. It operates over a wide range of collision energies, from $\sqrt{s_{\textrm{NN}}}$~=~3 GeV in fixed-target mode up to 200 GeV for ion collisions, and up to 510 GeV for proton-proton ($p+p$) collisions. This versatility is complemented by the STAR detector, which is equipped with multiple subsystems designed for measurements across an extensive kinematic range. A 3D model of the STAR experiment (from 2014) can be seen in Fig. \ref{STAR}.

The Time Projection Chamber (TPC), covering pseudorapidity |$\eta$|~<~1, provides precise tracking and momentum measurement of charged particles. It also serves as a tool for centrality determination through charged-particle multiplicity selection. The Barrel Electromagnetic Calorimeter (BEMC), also at |$\eta$|~<~1, detects energy deposits from photons, electrons and positrons, while additionally functioning as a fast online trigger. The Time-of-Flight (TOF) detector, operating within |$\eta$|~<~0.9, identifies particles based on their velocities and plays a role in mitigating pileup events.

At higher rapidities, the Vertex Position Detector (VPD) operates in the range 4.24~<~|$\eta$|~<~5.1, while the Zero Degree Calorimeter (ZDC) is positioned 18 meters downstream. Both serve as minimum bias triggers. Additionally, the VPD contributes to precise vertex reconstruction through its excellent timing capabilities, enhancing position resolution. The ZDC also plays a critical role in monitoring luminosity. Another detector consisting of two parts positioned far from the center of STAR is the Beam-Beam Counter (BBC), which operates in the range 3.4~<~|$\eta$|~<~5.0. It serves as a trigger for $p+p$ collisions and is also used to estimate a proxy for centrality in the $p$+Au collisions.

\section{Results}

In the analyses presented here, the jets are reconstructed using the anti-$k_{\mathrm{T}}$ algorithm. Two types of jets are considered: charged-particle jets, which are reconstructed solely from charged-particle tracks in the TPC, and full jets, which incorporate both charged-particle tracks from the TPC and energy deposits measured in the BEMC towers. Measurements from two collision systems are presented, $p+\mathrm{Au}$ and $\mathrm{Au}+\mathrm{Au}$ collisions at $\sqrt{s_{\mathrm{NN}}}~=~200$~GeV. The analyses benefit from a low constituent transverse momentum ($p_{\mathrm{T}}$) cut-off of 0.2~GeV/$c$, which allows for less biased jet reconstruction. Furthermore, the kinematic reach extends to jet transverse momenta of approximately 50--60~GeV/$c$ for the inclusive full jets, enabling meaningful comparisons with the results of the LHC experiments.

\subsection{Inclusive Charged-particle Jets in Au+Au Collisions}

In Au+Au collisions a strong effect of medium on the jet production is expected. One of the variables, which can be used to study effects of the medium is so called central-to-peripheral ratio $R_{\mathrm{CP}}$ defined as

\begin{equation}
R_{\mathrm{CP}}~=~\frac{\langle N_{\mathrm{coll}}^{\mathrm{per}} \rangle}{\langle N_{\mathrm{coll}}^{\mathrm{cent}} \rangle} \cdot
\frac{ \frac{1}{N_{\mathrm{evt}}^{\mathrm{AA, cent}}} \frac{\mathrm{d}^2 N^{\mathrm{jet}}_{\mathrm{AA, cent}}}{\mathrm{d} p_{T,\mathrm{jet}} \mathrm{d} \eta} }
{ \frac{1}{N_{\mathrm{evt}}^{\mathrm{AA, per}}} \frac{\mathrm{d}^2 N^{\mathrm{jet}}_{\mathrm{AA, per}}}{\mathrm{d} p_{T,\mathrm{jet}} \mathrm{d} \eta} },
\end{equation}
which compared jet yield measured in central collisions $\frac{\mathrm{d}^2 N^{\mathrm{jet}}_{\mathrm{AA, cent}}}{\mathrm{d} p_{T,\mathrm{jet}} \mathrm{d} \eta}$ scaled by the number of analyzed central events $N_{\mathrm{evt}}^{\mathrm{AA, cent}}$ to the jet yield measured in peripheral collisions $\frac{\mathrm{d}^2 N^{\mathrm{jet}}_{\mathrm{AA, per}}}{\mathrm{d} p_{T,\mathrm{jet}} \mathrm{d} \eta}$ scaled by the corresponding number of events $N_{\mathrm{evt}}^{\mathrm{AA, per}}$. The spectra are also scaled by the average number of binary collisions estimated from the Glauber model for central $\langle N_{\mathrm{coll}}^{\mathrm{cent}} \rangle$ and peripheral collisions $\langle N_{\mathrm{coll}}^{\mathrm{per}} \rangle$. The ratio differing from unity indicates modified spectra w.r.t. to simple superposition of $p+p$ collisions which may be due to jet quenching among other effects.

\begin{figure}[h]
\centering
\includegraphics[width=14 cm]{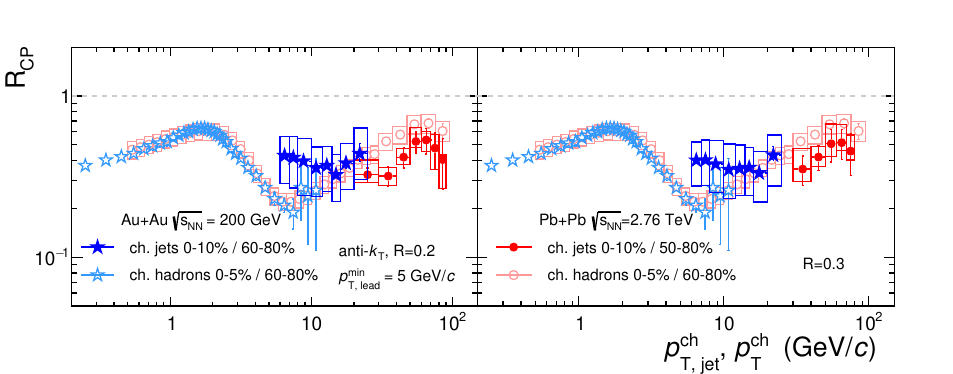}
\caption{The $R_{\mathrm{CP}}$ distributions for Au+Au collisions $\left(0-10\%/60-80\% \right)$ at $\sqrt{s_{\mathrm{NN}}}~=~200$~GeV compared to those measured in Pb+Pb collisions $\left(0-10\%/50-80\% \right)$ at $\sqrt{s_{\mathrm{NN}}}~=~2.76$~TeV~\cite{Abelev2014} for $R~=~0.2$ (left) and $R~=~0.3$ (right). Also shown are $R_{\mathrm{CP}}$ values for inclusive charged hadrons in Au+Au collisions $\left(0-5\%/60-80\% \right)$ at $\sqrt{s_{\mathrm{NN}}}~=~200$~GeV~\cite{Adams2003} and in Pb+Pb collisions $\left(0-5\%/60-80\% \right)$ at $\sqrt{s_{\mathrm{NN}}}~=~2.76$~TeV~\cite{Aad2015}. Data from RHIC are shown in blue, while data from the LHC are shown in red. Taken from Ref. \cite{Adam2020}.
\label{RCP}}
\end{figure}

Figure~\ref{RCP} presents a comparison of $R_{\mathrm{CP}}$ for charged-particle jets with $R = 0.2$ and $R = 0.3$ in Au+Au collisions at $\sqrt{s_{\mathrm{NN}}}~=~200$~GeV to measurements from Pb+Pb collisions at $\sqrt{s_{\mathrm{NN}}}~=~2.76$~TeV~\cite{Abelev2014}, as well as to $R_{\mathrm{CP}}$ for inclusive charged hadrons at both RHIC~\cite{Adams2003} and the LHC~\cite{Aad2015}. In central collisions, significant suppression of charged-particle jet yields is observed, reaching values around $R_{\mathrm{CP}} \sim 0.5$, largely independent of jet $p_{\mathrm{T}}$.

It is important to note, however, that the interpretation of these comparisons requires caution. The underlying spectral shapes, as well as the quark and gluon jet fractions, differ between RHIC and LHC energies, meaning that similar $R_{\mathrm{CP}}$ values do not imply identical quenching effects. Additionally, the $R_{\mathrm{CP}}$ for inclusive charged hadrons and that for jets cannot be directly compared at the same $p_{\mathrm{T}}$, as hadrons carry only a fraction of the parent jet's energy due to the fragmentation process.

The $R_{\mathrm{CP}}$ values for inclusive charged hadrons at RHIC and LHC are observed to be consistent within uncertainties over the overlapping $p_{\mathrm{T}}$ range. Similarly, the charged-particle jet $R_{\mathrm{CP}}$ shows comparable suppression at RHIC and LHC energies, though the $p_{\mathrm{T,jet}}^{\mathrm{ch}}$ ranges differ. The absence of a strong $p_{\mathrm{T}}$ dependence in jet $R_{\mathrm{CP}}$ contrasts with the more pronounced $p_{\mathrm{T}}$ dependence observed for charged hadrons, reflecting the distinct energy distribution between hadrons and jets. This comparison provides important constraints on theoretical models of jet quenching and highlights the need to carefully account for differences in partonic composition and fragmentation effects when interpreting suppression patterns across different collision systems and energies.

Similarly, as $R_{\mathrm{CP}}$ compares central and peripheral collisions, the nuclear modification factor $R_{\mathrm{AA}}$ is defined to compare collisions of nuclei to the collisions of protons.

\begin{equation}
R_{\mathrm{AA}}~=~\frac{1}{\langle N_{\mathrm{coll}} \rangle} \cdot 
\frac{\frac{1}{N_{\mathrm{evt}}^{\mathrm{AA}}} \frac{\mathrm{d}^2 N^{\mathrm{jet}}_{\mathrm{AA}}}{\mathrm{d} p_{\mathrm{T,jet}} \mathrm{d}\eta}}
{\frac{1}{N_{\mathrm{evt}}^{pp}} \frac{\mathrm{d}^2 N^{\mathrm{jet}}_{pp}}{\mathrm{d} p_{\mathrm{T,jet}} \mathrm{d}\eta}},
\end{equation}
where $\frac{\mathrm{d}^2 N^{\mathrm{jet}}_{\mathrm{AA}}}{\mathrm{d} p_{\mathrm{T,jet}} \mathrm{d}\eta}$ is jet yield in the nucleus+nucleus collision, scaled by the number of analysed events $N_{\mathrm{evt}}^{\mathrm{AA}}$, $\frac{\mathrm{d}^2 N^{\mathrm{jet}}_{pp}}{\mathrm{d} p_{\mathrm{T,jet}} \mathrm{d}\eta}$ is the yield measured in the $p+p$ collisions, scaled by the corresponding events $N_{\mathrm{evt}}^{pp}$. The $p+p$ spectra are scaled up by the average number of binary collisions $\langle N_{\mathrm{coll}} \rangle$ for the heavy-ion collisions calculated from the Glauber model.

\begin{figure}[h]
\centering
\includegraphics[width=14 cm]{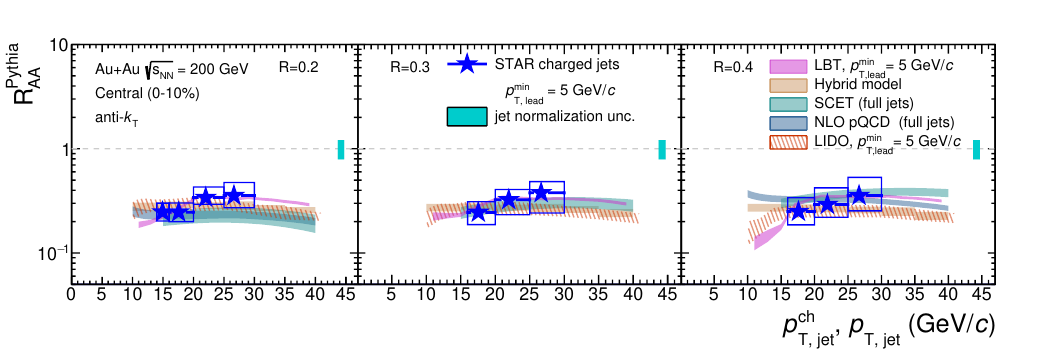}
\caption{Comparison of $R_{\mathrm{AA}}$ (stars) measured in the Au+Au collisions at $\sqrt{s_{\mathrm{NN}}}$~=~200 GeV to the theoretical calculations. The Hybrid \cite{CasalderreySolana2019}, LBT \cite{He2019}, and LIDO \cite{Ke2019, He2015} calculations correspond to charged-particle jets, while SCET \cite{Chien20161, Chien20162} and the NLO \cite{Vitev2010} calculations refer to fully reconstructed jets. Taken from Ref. \cite{Adam2020}. \label{RAA}}
\end{figure}

Figure \ref{RAA} presents a comparison of the measured charged-particle jet $R_{\mathrm{AA}}$ to various theoretical calculations. The reference distribution is given by the inclusive charged-particle jet spectrum in $p+p$ collisions at $\sqrt{s}~=~200$~GeV, as calculated using \textsc{Pythia} Monte Carlo generator version 6.428 \cite{Sjoestrand20060501}, with the Perugia 2012 tune (370) and CTEQ6L1 LO parton distribution functions \cite{Skands2010}. The Hybrid \cite{CasalderreySolana2019}, LBT \cite{He2019}, and LIDO \cite{Ke2019, He2015} models provide predictions for charged-particle jets, while the SCET \cite{Chien20161, Chien20162} and NLO pQCD calculations \cite{Vitev2010} correspond to fully reconstructed jets. Since the $p_{\mathrm{T,jet}}$ dependence of full jet $R_{\mathrm{AA}}$ is weak, a comparison between these calculations and the charged-particle jet data remains meaningful. Additionally, the LBT and LIDO models incorporate a cut on the leading constituent in the Au+Au spectrum, consistent with the $p_{\mathrm{T,lead}}^{\mathrm{min}}~=~5$~GeV/$c$ requirement applied in the experimental analysis. All theoretical predictions are found to be consistent with the measured inclusive jet $R_{\mathrm{AA}}$ within uncertainties in the unbiased kinematic region. The most significant differences among models appear for $R~=~0.4$, indicating that ongoing measurements of inclusive jet $R_{\mathrm{AA}}$ may have the potential to distinguish between these theoretical approaches.

\subsection{Semi-inclusive Recoil Jet Yield Modification}

A powerful tool to investigate the properties of the QGP is the study of jets recoiling from high transverse energy ($E_{\mathrm{T}}$) direct photons ($\gamma_{\mathrm{dir}}$)~\cite{Wang1996}. Since $\gamma_{\mathrm{dir}}$ photons do not undergo strong interactions with the medium, they provide a clean probe of jet quenching effects. Furthermore, comparing measurements of $\gamma_{\mathrm{dir}}$+jet and $\pi^{0}$+jet events can offer insights into the color factor and path-length dependence of energy loss mechanisms~\cite{Adamczyk2016}. The study of recoil jet distributions for different jet cone radii also serves as a sensitive probe of in-medium jet broadening.

In this section, an analysis of fully-corrected, semi-inclusive distributions of charged-particle jets recoiling from high-$E_{\mathrm{T}}$ $\gamma_{\mathrm{dir}}$ and $\pi^{0}$ triggers in central Au+Au collisions at $\sqrt{s_{\mathrm{NN}}}~=~200$~GeV is reported. The dataset was collected during the 2014 RHIC run, utilizing a trigger condition that required an energy deposition greater than 5.6~GeV in a single tower of the STAR Barrel Electromagnetic Calorimeter (BEMC).

A variable similar to the nuclear modification factor is defined for the semi-inclusive collisions. This variable $I_{\mathrm{AA}}$ is defined as

\begin{equation}
    I_{\mathrm{AA}}~=~\frac{Y^{\mathrm{A+A}}\left( p_{\mathrm{T,jet}}^{\mathrm{ch}}, R \right)}{Y^{p+p}\left( p_{\mathrm{T,jet}}^{\mathrm{ch}}, R \right)},
\end{equation}
and compares trigger-normalized jet yields in recoil acceptance for nucleus + nucleus collisions $Y^{\mathrm{A+A}}\left( p_{\mathrm{T,jet}}^{\mathrm{ch}}, R \right)$ and $p+p$ collisions $Y^{p+p}\left( p_{\mathrm{T,jet}}^{\mathrm{ch}}, R \right)$.

\begin{figure}[h]
\centering
\includegraphics[width=8 cm]{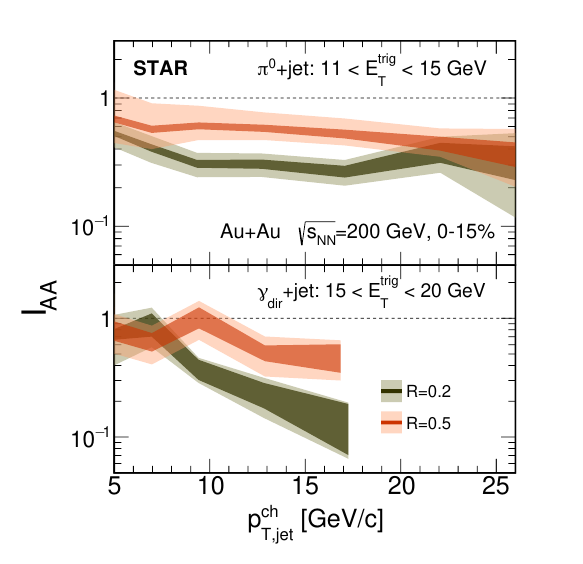}
\caption{The $I_{\mathrm{AA}}$ of semi-inclusive recoil jet distributions in central Au+Au collisions at $\sqrt{s_{\mathrm{NN}}}$~=~200 GeV shown for both $\gamma_{\mathrm{dir}}$ and $\pi^{0}$ triggers, with $R~=~0.2$ and $R~=~0.5$. The uncertainty bands reflect the correlated uncertainties in both the numerator and denominator. The dark bands indicate statistical uncertainties, while the light bands represent systematic uncertainties. Taken from \cite{STAR2023}.
\label{IAA}}
\end{figure}  

Figure \ref{IAA} presents the $I_{\mathrm{AA}}$ observable, comparing jet yield distributions in central (0-15~\%) Au+Au collisions to those of $p+p$ collisions, for a common trigger selection and jet resolution parameter $R$, shown for the highest measured $E_{\mathrm{T}}^{\mathrm{trig}}$ bin for each trigger type (11~<~$E_{\mathrm{T}}^{\mathrm{trig}}$~<~15 GeV for $\pi^0$+jet and 15~<~$E_{\mathrm{T}}^{\mathrm{trig}}$~<~20 GeV for $\gamma_{\mathrm{dir}}$+jet). A significant suppression ($I_{\mathrm{AA}} < 1$) of recoil jet yields is observed in central Au+Au collisions for $R = 0.2$, while the suppression is noticeably reduced for $R = 0.5$. This suppression arises from both the $p_{\mathrm{T}}$ spectrum shape effects~\cite{STAR2023} and parton energy loss in the medium. The latter reflects population-averaged parton energy loss, where the "population" consists of recoil jets initiated by partons with varying energies, flavors, and in-medium path lengths. In semi-inclusive hadron-triggered events, trigger bias favors hadrons produced near the QGP surface, causing the recoil jets to traverse longer distances through the medium. This geometric bias leads to an averaged suppression over diverse partonic trajectories. The reduced suppression observed at larger $R$ values suggests that a greater fraction of the jet’s energy is recovered within the jet cone, providing access to the angular scale of medium-induced energy redistribution.
 
$I_{\mathrm{AA}}$ is consistent within uncertainties for recoil jets triggered by both direct photons ($\gamma_{\mathrm{dir}}$) and $\pi^{0}$ mesons within the same $E_{\mathrm{T}}^{\mathrm{trig}}$ bin, despite the steeper spectrum associated with $\pi^{0}$ triggers. This observation indicates a larger average medium-induced energy loss for recoil jets associated with $\pi^{0}$ triggers, offering new constraints on the flavor and path-length dependence of jet quenching mechanisms~\cite{STAR2023}.

\subsection{Semi-inclusive Yield Modification in $p$+Au Collisions}

In contrast to Au+Au collisions, where event activity (EA) or centrality is typically classified based on charged particle multiplicity at midrapidity~\cite{Miller20071101}, defining a centrality proxy in small systems like $p+\mathrm{Au}$ is more challenging due to significant contributions from hard scatterings, which introduce autocorrelations between jet production and the EA measurement. To mitigate this, a large separation in pseudorapidity between the EA measurement and midrapidity jets is required.

\begin{figure}[h]
\centering
\includegraphics[width=12 cm]{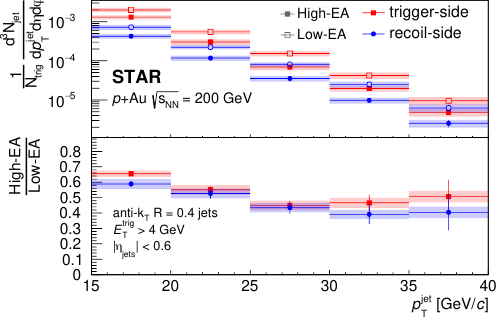}
\caption{Jet spectra per trigger for the trigger-side ($|\phi_{\mathrm{jet}} - \phi_{\mathrm{trig}}|~<~\pi/3$) and recoil-side ($|\phi_{\mathrm{jet}} - \phi_{\mathrm{trig}}|~>~2\pi/3$) jets measured in $p$+Au collisions at $\sqrt{s_{\mathrm{NN}}}~=~200$~GeV are shown in the top panel. Jets are reconstructed with $R~=~0.4$, and the offline trigger requirement is $E_{\mathrm{T}}^{\mathrm{trig}}~>~4$~GeV. Spectra are presented for both high-EA and low-EA event classes. The bottom panel shows the ratio of semi-inclusive jet spectra in high-EA to low-EA events. Statistical uncertainties are represented by the error bars, while systematic uncertainties are indicated by the shaded boxes. Taken from \cite{Abdulhamid2024}. \label{pAu}}
\end{figure}

In this analysis, STAR utilizes the Beam-Beam Counters located in the Au-going direction, covering $\eta_{\mathrm{BBC}}~\in~[-5, -3.4]$, to quantify EA based on the sum of ADC signals from scintillating tiles. This method provides a wide rapidity gap relative to jets measured at $|\eta| \leq 0.6$, effectively reducing autocorrelation effects. The EA distribution measured by the BBC is used to classify events into low- and high-EA categories, defined respectively as the lowest and highest $30\%$ of the minimum bias EA distribution. Despite detector signal saturation effects, correlations between EA measured by the BBC and midrapidity charged particle density confirm the suitability of this method as a centrality proxy in $p+\mathrm{Au}$ collisions.   

Figure~\ref{pAu} presents the first fully-corrected semi-inclusive jet spectra in small system collisions at RHIC. The jet spectra per trigger are shown for both trigger-side and recoil-side jets, with an azimuthal selection of $\Delta\phi~<~\pi/3$ around the trigger direction or opposite to it for recoil jets. A clear suppression of both trigger- and recoil-side jet yields is observed in high-EA events compared to low-EA events. Notably, the suppression is comparable for both sides, which differs from typical jet quenching signatures in large systems, where recoil-side jets are more strongly suppressed. This suggests that the observed suppression in small systems may arise from anti-correlation between event activity and the underlying hard scattering scale, rather than genuine in-medium energy-loss effects.

\section{Conclusions}

Jets serve as essential probes for studying in-medium energy loss in nucleus-nucleus collisions. The STAR experiment observes a significant suppression of jet yields in $\mathrm{Au}+\mathrm{Au}$ collisions at $\sqrt{s_{\mathrm{NN}}} = 200$~GeV, consistent with strong interactions of jets with the quark-gluon plasma. In $p+\mathrm{Au}$ collisions, a suppression is also observed; however, its characteristics are inconsistent with expectations from in-medium energy loss and are instead attributed to a negative correlation between event activity and the underlying hard scale $Q^2$. Future measurements with larger datasets will provide improved precision, extended kinematic reach, and enhanced capabilities for full jet reconstruction.

\textbf{Acknowledgments}: The work has been supported by the Czech Science Foundation grant 23-07499S.

\bibliographystyle{unsrt}  


\end{document}